\documentclass[12pt]{article}
\usepackage{epsf}
\usepackage{amsmath}

\usepackage{graphics}
\usepackage{cite}

\begin{titlepage}

\begin{document}

\hfill April 2024

\begin{center}

{\bf \LARGE Status of Electromagnetic Accelerating Universe}
\footnote{Submitted to {\it Particle Theory and Theoretical Cosmology},
Festschrift for 80th anniversary of P.H. Frampton. To appear in {\it Entropy}
journal, and Special Edition, MDPI Books (2024) }\\
\vspace{2.5cm}
{\bf Paul H. Frampton}\footnote{paul.h.frampton@gmail.com}\\
\vspace{0.5cm}
{\it Dipartimento di Matematica e Fisica "Ennio De Giorgi",\\ 
Universit\`{a} del Salento and INFN-Lecce,\\ Via Arnesano, 73100 Lecce, Italy.
}

\vspace{1.0in}
\end{center}

\begin{abstract}
\noindent
\end{abstract}
\noindent
To describe the dark side of the Universe, we adopt a novel
approach where dark energy is explained as an electrically charged majority of
dark matter. Dark energy, as such, does not exist. The Friedmann equation
at the present time coincides with that in a conventional approach, although
the cosmological "constant" in the Electromagnetic Accelerating Universe (EAU) Model shares a time dependence with the matter
component. Its equation of state is $\omega \equiv P/\rho \equiv -1$
within observational accuracy.

\end{titlepage}

\noindent
\section{Introduction to the EAU Model} 
\bigskip

\noindent
Theoretical cosmology is at an exciting stage because about 95\% of the energy in
the Visible Universe remains incompletely understood. The 25\% which
is Dark Matter has constituents whose mass is unknown by over one hundred
orders of magnitude. The 70\% which is Dark Energy is, if anything, more mysterious: 
although it can be parametrised by a Cosmological Constant with equation of state
$\omega = -1$ which provides an excellent phenomenological description,
that is only a parametrisation and not a complete understanding.

\bigskip

\noindent
in the present paper, we address the issues of Dark Matter and Dark Energy
using a novel approach. We use only the classical theories of
electrodynamics and general relativity.
We shall not employ any knowledge of quantum mechanics or of
theories describing the short range strong and weak interactions.

\bigskip

\noindent
This paper may be regarded as a follow up to our 2018 paper
\cite{Frampton2018}  entitled {\it On the Origin and Nature of Dark
Matter} and could have simply added {\it and Energy}
to that title. We have, however, chosen {\it Status of Electromagnetic
Accelerating Universe} because it more accurately characterises
our present emphasis on the EAU model whose main idea that electromagnetism
dominates over gravitation in the explanation of the accelerating
cosmological expansion. This idea takes us beyond the first paper\cite{Einstein}
which applied general relativity to theoretical cosmology.
This is not surprising, since in 1917 that author was obviously unaware
of the fact \cite{Perlmutter,Riess} discovered only in 1998 that the rate of cosmological expansion
is accelerating.

\bigskip

\noindent
The make up of this paper is that Primordial Black Holes
are discussed  in Section 2, then Primordial Naked
Singularities in Section 3. Section 4 contains a possible
supporting evidence for the EAU model based on the
recently reported Amaterasu cosmic ray. Finally in 
Section 5 there is a Discussion.

\newpage

\section{Primordial Black Holes (PBHs)}

\noindent
Black holes in the universe fall into two classes, those which arise from the
gravitational collapse of stars and others which do not. By primordial, we
refer to all of the others. In general, PBHs with masses up to $10^5 M_{\odot}$
are expected to be formed during the first second after the Big Bang and
arise from inhomogeneities and fluctuations of spacetime. The existence
of PBHs was first posited in Russia \cite{Novikov} by Novikov and Zeldovich
and independently seven years later in the West by Carr and Hawking\cite{CarrHawking}.
The idea that the dark matter
constituents are PBHs was first suggested by Chapline\cite{Chapline}.

\bigskip

\noindent
Shortly after the original presentation of general relativity\cite{Einstein1915}
a metric describing a static black hole of mass M with zero charge
and zero spin was discovered by Schwarzschild\cite{Schwarzschild1916}
in the form
\begin{equation}
ds^2 = - \left( 1 - \frac{r_S}{r} \right) dt^2 + \left( 1 - \frac{r_S}{r} \right)^{-1} dr^2 + r^2 d\Omega^2
\label{Schwarzschild}
\end{equation}
\noindent
Shortly thereafter, the Reissner-Nordstrom metric\cite{Reissner}
for a static Black Hole with electric charge was found. It then took 
a surprising forty-five years
until Kerr cleverly found a solution\cite{Kerr1963} of general relativity
corresponding to a such a solution with spin. We shall not
discuss the case of non-zero spin in the present paper because,
although we expect that all the objects we discuss do spin in
Nature, according to the calculations in \cite{KNEA} which
use Kerr's generalisation, spin is an inessential 
complication in all of our subsequent considerations.

\bigskip

\subsection{Primordial Intermediate Mass Black Holes \\
(PIMBHs) as Galactic Dark Matter}

\noindent
According to global analyses of the cosmological parameters
about one quarter of the energy of the universe is in the form of 
electrically-neutral dark matter. It has been proposed
\cite{Frampton} that the dark matter constituents are black holes with masses many
times the mass of the Sun. In a galaxy like the Milky Way, the proposal is that
residing in the galaxy are between ten million and ten billion black holes with masses between
one hundred and one hundred thousand solar masses.

\bigskip

\noindent
Black holes in this range of masses are commonly known as Intermediate Mass Black Hole (IMBHs)
since they lie above the masses of stellar-mass black holes
and below the masses of the
supermassive black holes at galactic centres. It has long been mysterious
why there is a mass gap between stellar-mass and supermassive black holes.
If the proposed solution of the galactic dark matter problem is correct, it will answer this old
question.

\bigskip

\noindent
There is irrefutable evidence for stellar-mass black holes from observations of X-ray binaries.
Such systems were first emphasized in \cite{Guseynov} then further studied in \cite{Trimble}. All
the known stellar-mass black holes are members of X-ray binaries. The first was
discovered sixty years
ago in 1964 in Cygnus X-1 and many stellar-mass black holes have since been discovered
from studies of X-ray binaries, with masses in a range between 
$5M_{\odot}$ and  $100M_{\odot}$,
where the first-discovered Cygnus X-1 is at about $15M_{\odot}$.

\bigskip

\noindent
We note historically that dark matter was first discovered by 
Zwicky\cite{Zwicky,Zwicky2} in 1933
in the Coma Cluster, and its presence in galaxies was demonstrated
convincingly by Rubin in the 1960s and 1970s 
from the rotation curves of many galaxies 
\cite{Rubin}.  

\bigskip

\noindent
Regarding the PBH mass range, the possible PBH masses extend upwards to many solar
masses and above, far beyond
what was was thought possible twenty years ago when ignorance about
PBHs with many solar masses probably prevented the MACHO
\cite{MACHO} and EROS\cite{EROS} Collaborations from 
discovering more of the dark matter.

\bigskip

\noindent
If all black holes were formed by gravitational collapse then black holes with
$M_{BH} \ll M_{\odot}$ would be impossible because stars powered by nuclear fusion
cannot be far below $M = M_{\odot}$.
It was first suggested in \cite{Novikov,CarrHawking} 
that black holes can be produced in the early stages of the cosmological expansion..

\bigskip

\noindent
Such PBHs are of special interest for several reasons. Firstly, they are the only type
of black hole which can be so light, down to $10^{12}kg\sim10^{-18}M_{\odot}$, that Hawking radiation might conceivably be detected. Secondly, PBHs in the intermediate-mass region
$100M_{\odot} \leq M_{IMBH} \leq 10^5M_{\odot}$ can provide the galactic
dark matter. 

\bigskip

\noindent
The mechanism of PBH formation involves large fluctuations or inhomogeneities. 
Carr and Hawking\cite{CarrHawking}
argued that we know there are fluctuations in the universe in order to seed
structure formation and there must similarly be fluctuations in the early universe. Provided the
radiation is compressed to a high enough density, meaning to a radius as small
as its Schwarzschild radius, a PBH will form. Because the density in the early
universe is extremely high, it is very likely that PBHs will be created. The two
necessities are high density which is guaranteed and large inhomogeneities.

\bigskip

\noindent
During radiation domination
\begin{equation}
a(t) \propto t^{1/2}
\label{raddom}
\end{equation}
and 
\begin{equation}
\rho_{\gamma} \propto a(t)^{-4} \propto t^{-2}
\label{rhogamma}
\end{equation}

\bigskip

\noindent
Ignoring factors $O(1)$ and bearing in
mind that the radius of a black hole is 
\begin{equation}
r_{BH} \sim \left( \frac{M_{BH}}{M_{Planck}^2} \right)
\label{BHradius}
\end{equation}
with
\begin{equation}
M_{Planck} \sim 10^{19}GeV \sim 10^{-8} kg \sim 10^{-38}M_{\odot}
\label{MPlanck}
\end{equation} 
and using the Planck density $\rho_{Planck}$ 
\begin{equation}
\rho_{Planck} \equiv (M_{Planck})^4 \sim (10^{-5}g)(10^{-33}cm)^{-3} = 10^{94} \rho_{H_2 O}
\label{rhoPlanck} 
\end{equation}
the density of a general black hole $\rho_{BH}(M_{BH})$ is
\begin{equation}
\rho_{BH}(M_{BH}) \sim \left( \frac{M_{BH}}{r_{BH}^3} \right) = \rho_{Planck} \left(
\frac{M_{Planck}}{M_{BH}} \right)^2 \sim 10^{94} \rho_{H_2 O} \left( \frac{10^{-38} M_{\odot}}{M_{BH}} \right)^2
\label{rhoBH}
\end{equation}
which means that for a solar-mass black hole
\begin{equation}
\rho_{BH}(M_{\odot}) \sim 10^{18} \rho_{H_2 O}
\label{rhoBHSun}
\end{equation}
while for a billion solar mass black hole
\begin{equation}
\rho_{BH}(10^9M_{\odot}) \sim \rho_{H_2 O}.
\label{rhoBHSM}
\end{equation}
and above this mass the density falls as $M_{BH}^{-2}$.

\bigskip

\noindent
The mass of the PBH is derived by combining
Eqs. (\ref{rhogamma}) and (\ref{rhoBH}). We see from these 
two equations that $M_{PBH}$ grows linearly with time and using Planckian units 
or Solar units we find respectively
\begin{equation}
M_{PBH} \sim \left( \frac{t}{10^{-43} sec} \right) M_{Planck}
\sim \left( \frac{t}{1 sec} \right) 10^5 M_{\odot}
\label{PBHmass}
\end{equation}
which implies, if we insisted on PBH formation
before the electroweak phase transition, $t < 10^{-12}s$, that
\begin{equation}
M_{PBH} < 10^{-7} M_{\odot}
\label{upperbound}
\end{equation}

\bigskip

\noindent
The upper bound in Eq.(\ref{upperbound}) explains historically why the MACHO
searches around 2000 \cite{MACHO,EROS}, inspired by the 1986
suggestion of Paczynski\cite{Paczynski}, lacked motivation
to pursue searching beyond $100M_{\odot}$ because it was thought 
incorrectly at that
time that PBHs were too light. It was known correctly
that the results of gravitational collapse of
normal stars, or even large early stars, were below $100M_{\odot}$.
Supermassive black holes with $M > 10^6M_{\odot}$
such as $Sag A^*$ in the Milky Way were beginning to be discovered in galactic centers
but their origin was unclear and will be
discussed further in the following Section 2.2. 

\bigskip

\noindent
Hawking radiation implies that the lifetime for a black hole evaporating 
{\it in vacuo} is given by the cubic formula
\begin{equation}
\tau_{BH} \sim \left( \frac{M_{BH}}{M_{\odot}} \right)^3 \times 10^{64} years
\label{BHlifetime}
\end{equation}
so that to survive for the age $10^{10}$ years of the universe, there is a
lower bound on $M_{PBH}$ to augment the upper bound in Eq.(\ref{upperbound}),
giving as the full range of Carr-Hawking PBHs:
\begin{equation}
10^{-18}M_{\odot} < M_{PBH} < 10^{-7} M_{\odot}
\label{CarrHawking}
\end{equation}
The lowest mass surviving PBH in Eq.(\ref{CarrHawking}) has the density
$\rho \sim 10^{58} \rho_{H_2 O}$. It is an awesome object which has the physical size of a proton and
the mass of Mount Everest.

\bigskip

\noindent
The Hawking temperature $T_H(M_{BH})$
of a black hole is
\begin{equation}
T_H(M_{BH}) = 6 \times 10^{-8}K \left( \frac{M_{\odot}}{M_{BH}} \right)
\label{Hawking}
\end{equation}
which would be above the CMB temperature, and hence there would be outgoing radiation for all of the cases with $M_{BH} < 2 \times 10^{-8}M_{\odot}$.
Hypothetically, if the dark matter halo were made entirely of the brightest possible (in terms of Hawking radiation) $10^{-18}M_{\odot}$
PBHs, the expected distance to the nearest PBH would be about $10^7$ km.
Although the PBH temperature, according to Eq. (\ref{Hawking}) is $\sim 6\times 10^{10} K$,
the inverse square law 
renders the intensity of Hawking radiation too small, by many orders of magnitude,
to allow its detection by any foreseeable apparatus on Earth.

\bigskip

\noindent
The original mechanism produces PBHs with masses in the range
up to $10^{-7} M_{\odot}$. We shall now discuss formation of far more
massive PBHs
by a different mechanism.
As discussed, PBH formation requires very large inhomogeneities.
Here we shall briefly illustrate how mathematically to produce inhomogeneities
which are exponentially large.

\bigskip

\noindent
In a single inflation, no exceptionally large density perturbation is expected. Therefore
we use two-stage hybrid inflation with respective fields called \cite{LiddleLyth},
inflaton and waterfall. The idea of parametric
resonance is that after the first inflation mutual couplings of the
inflaton and waterfall fields cause both to oscillate wildly
and produce perturbations which grow exponentially.
The secondary (waterfall) inflation then stretches further these inhomogeneities,
enabling production of PBHs with arbitrarily high mass.
The specific model provides an existence theorem to confirm that
arbitrary mass PBHs can be produced. The resulting mass function
is spiked, but it is possible that other PBH production mechanisms can
produce a smoother mass function, as deserves further study.
The details of the model are in \cite{FKTY} where the inflaton
and waterfall fields are denoted by $\sigma$ and $\psi$ respectively.

\noindent
Between the two stages of inflation, the $\sigma$ and $\psi$ fields oscillate, decaying into their quanta via their self and mutual couplings. Specific modes of $\sigma$ and $\psi$ are amplified by
parametric resonance. The resulting equation which couples the 
two fields is of Mathieu type 
with the required exponentially-growing solutions. Numerical
solution shows that the peak wave number $k_{peak}$ is approximately linear in
$m_{\sigma}$. The resultant PBH mass, the horizon 
mass when the fluctuations re-enter the horizon, is approximately
\begin{equation}
M_{PBH} \sim 1.4 \times 10^{13}M_{\odot} \left( \frac{k_{peak}}{Mpc^{-1}} \right)^{-2}
\label{PBHmass}
\end{equation}
Explicit plots were exhibited in \cite{FKTY} 
for the cases $M_{PBH} = 10^{-8}M_{\odot},
 10^{-7}M_{\odot}$ and $10^5M_{\odot}$
but it was checked at that time in 2010 that the parameters 
can be chosen to produce arbitrarily high
PBH mass. 

\bigskip

\noindent
In this production mechanism based on hybrid inflation with parametric 
resonance, the mass function is sharply spiked at a specific mass region.
Whether such a mass function is a general feature of PBH formation, or
is only a property of this specific mechanism, merits further study.
The mechanism demonstrates the possibility of primordial formation of black holes
with many solar masses. For completeness, it should be pointed out that PBHs with
masses up to $10^{-15}M_{\odot}$ were discussed already in the 1970s,
for example by Carr\cite{Carr1975} and by Novikov, Polnarev, Starobinskii
and Zeldovich\cite{NPSZ1979}.

\bigskip

\noindent
For dark matter in galaxies, PIMBHs are important,
 where the upper end may be truncated at 
$10^{5}M_{\odot}$ to stay well away 
from galactic disk instability \cite{Ostriker}. 

\bigskip

\noindent
The dark matter in the Milky Way fills out an approximately spherical halo
somewhat larger in radius than the disk occupied by the luminous stars.
Numerical simulations of structure formation suggest a profile of the
dark matter of the NFW types\cite{NFW}. The NFW profile is 
independent
of the mass of the dark matter constituent.

\bigskip

\noindent
Our discussion\cite{Frampton} focused on galaxies like the Milky Way
and restricted the mass range for the appropriate dark matter 
to only three orders of magnitude
\begin{equation}
10^2 M_{\odot} < M < 10^5M_{\odot}
\label{Frampton}
\end{equation}

\bigskip

\noindent
We shall not repeat the arguments here, just to say that the
constituents are Primordial Intermediate Mass Black Holes, PIMBHs. Given a
total dark halo mass of $10^{12}M_{\odot}$, the number $N$ of PIMBHs
is between ten million ($10^7$) and ten billion ($10^{10}$) 
Assuming the dark halo has radius $R$ of a hundred thousand ($10^5$)
light years
the mean separation $\bar{L}$ of PIMBHs can be estimated by
\begin{equation}
\bar{L} \sim \left( \frac{R}{N} \right)
\label{meanL}
\end{equation}
which translates to
\begin{equation}
100 ly < \bar{L} < 1000 ly
\label{meanL2}
\end{equation}
which is also an estimate of the distance of the nearest PIMBH
to the Earth.

\bigskip

\noindent
It may be surprising that as many as $10^7 \leq N \leq 10^{10}$ intermediate-mass black holes
in the Milky Way have remained undetected. They could have been
detected more than two decades ago had the MACHO Collaboration
\cite{MACHO} persisted in its microlensing experiment at Mount Stromlo
Observatory in Australia.

\bigskip

\noindent
The first discovery of dark matter by Zwicky \cite{Zwicky,Zwicky2} 
was in the Coma cluster which is a large cluster at 99 Mpc
containing over a thousand galaxies and with
total mass estimated at $6 \times 10^{14}M_{\odot}$ \cite{Merritt}.
A nearer cluster at 16.5 Mpc is the Virgo cluster with over two thousand
galaxies and whose mass $\sim 10^{15}M_{\odot}$
is also dominated by dark matter, as well as a small
amount of X-ray emitting gas \cite{Virgo1,Virgo2}.
A proof of the existence (if more were needed)
of cluster dark matter was provided by the Bullet cluster
collision where the distinct behaviors of the X-ray emitting gas which
collides, and the dark matter which does not collide, was clearly observable
\cite{Bullet}.

\bigskip

\noindent
Since there is not the same disk stability limit as for galaxies,
the constituents of the cluster dark matter can involve also PSMBHs 
up to much higher masses.

\bigskip

\noindent
Such a solution of the galactic dark matter problem cries out for experimental
verification. Three methods have been discussed: wide binaries,
distortion of the CMB, and microlensing. Of these, microlensing seems the most
direct and the most promising.

\bigskip

\noindent
Microlensing experiments were carried out by the MACHO\cite{MACHO}
and EROS\cite{EROS} Collaborations many years ago. At that time, it was
believed that PBH masses were below $10^{-7}M_{\odot}$ by virtue of the
Carr-Hawking mechanism. Heavier black holes could, it was then believed,
arise only from gravitational collapse of normal stars, or 
heavier early stars, and would have mass below $100M_{\odot}$. 

\bigskip

\noindent
For this reason, there was no motivation to suspect that there might be
MACHOs which led to higher-longevity microlensing events.
The longevity, $\hat{t}$, of an event is
\begin{equation}
\hat{t} = 0.2 yrs \left( \frac{M_{PBH}}{M_{\odot}} \right)^{\frac{1}{2}}
\label{longevity}
\end{equation}
which assumes a transit velocity $200 km/s$.
Subsituting our extended PBH masses, one finds approximately
$\hat{t} \sim 6, 20, 60$ years for $M_{PBH} \sim 10^3, 10^4, 10^5 M_{\odot}$
respectively, and searching for light curves with these higher
values of $\hat{t}$ could be very rewarding.

\bigskip

\noindent
Our understanding is that the original telescope used by the
MACHO Collaboration\cite{MACHO} at the Mount Stromlo
Observatory in Australia was accidentally
destroyed by fire, and that some
other appropriate telescopes are presently being used to search
for extasolar planets, of which over six thousand are already known. 
It is seriously hoped that MACHO searches will resume and
focus on greater longevity microlensing events. Some encouragement
can be derived from this, by a member of the 
original MACHO Collaboration 
{\it There is no known problem with searching for events of greater longevity than those discovered in 2000; only the longevity of the people!}
It is possible that convincing observations showing only a fraction of the light curves
could suffice? If so, only a fraction of the {\it e.g.} six years, corresponding to PIMBHs with
one thousand solar masses, could be enough to confirm the theory.

\bigskip

\subsection{Primordial Supermassive Black Holes \\
(PSMBHs) at Galactic Centers}

\bigskip

\noindent
There is observational evidence for supermassive black holes from the
observations of fast-moving stars around them and such stars being swallowed or 
torn apart by the strong gravitational field. The first discovered SMBH was naturally
the one, Sag $A^*$, at the core of the Milky Way which was discovered
in 1974 and has mass $M_{SagA*} \sim 4.1 \times 10^6M_{\odot}$.
SMBHs discovered at galactic cores include those for galaxies named M31,
NGC4889, among many others. The SMBH at the core of 
the nearby Andromeda galaxy ($M31$)
has mass $M=2\times 10^8M_{\odot}$, fifty times $M_{SagA*}$.
The most massive core SMBH so far observed is for NGC4889 with 
$M \sim 2.1\times 10^9M_{\odot}$. Some galaxies contain two SMBHs in a binary, believed to be the result of a galaxy merger. 
Quasars contain black holes with even higher masses up to at least
$4\times 10^{10}M_{\odot}$. .

\bigskip

\noindent
A black hole with the mass of $Sag A^*$ would disrupt the disk
dynamics \cite{Ostriker} were it out in the spiral arms but at, or
near to, the
center of mass it is more stable. $Sag A^*$ is far too massive to
have been the result of a gravitational collapse, and if we take
the view that all black holes either are 
the result of gravitational collapse or are primordial then the
galaxies' core SMBHs must be primordial.

\bigskip

\noindent
Nevertheless, it is probable that the PSMBHs are
built up by merging and accretion from less
massive PIMBH seeds.

\newpage

\section{Primordial Naked Singularities (PNSs)}

\noindent
Just as neutral black holes can be formed as PBHs in the early
universe, we expect that objects can be formed based on the
Reissner-Nordstrom metric \cite{Reissner}
\begin{equation}
ds^2 = f(r) dt^2 - f(r)^{-1}dr^2 - r^2 d\theta^2 -r^2 \sin^2 \theta d\phi^2
\label{RNmetric}
\end{equation}
where
\begin{equation}
f(r) \equiv \left( 1 - \frac{r_S}{r} + \frac{r_Q^2}{r^2} \right).
\label{f(r)}
\end{equation}
with
\begin{equation}
r_S =   2 G M ~~~~~~ r_Q= Q^2 G
\label{rSrQ}
\end{equation}

\bigskip

\noindent
The horizon(s) of the RN metric occur when
\begin{equation}
f(r) =0
\end{equation}
which gives
\begin{equation}
r_{\pm} = \frac{1}{2} \left( r_S \pm \sqrt{r_S^2 - 4 r_Q^2} \right)
\label{rpm}
\end{equation}

\bigskip

\noindent
For $2r_Q < r_S$, $Q^2 < M$, there are two horizons. When $2r_Q = r_S$,
$Q^2=M$ 
the RN black hole is extremal and there is only one horizon.
If $2r_Q > r_S$,  $Q^2 > M$, the  RN metric is super-extremal, there is no horizon
at all  and the $r=0$ singularity is observable to a distant observer. This is known as a naked
singularity and with this last inequality it is no longer a black hole which, by definition, requires an horizon.

\bigskip

\noindent
Consider two identical objects with mass M and charge Q. Then
the electromagnetic repulsive force $F_{em} \propto k_eQ^2$
and the gravitational attraction $F_{grav} \propto GM^2$.
Thus, for the electromagnetic repulsion to exceed the gravitational
attraction we need $Q^2 > GM^2/k_e$ and hence
perhaps super-extremal Reissner-Nordstrom or Naked
Singularities(NSs)\footnote{To anticipate NSs we shall replace
BH by NS for charged dark matter. If charges satisfy $Q^2<M$
this replacement is unnecessary.} We cannot claim to understand the formation
of PNSs. One idea hinted at in \cite{Araya2022} is that extremely massive
ones, charger PEMNSs might begin life as electrically neutral PBHs
which selectively accrete electrons over protons. However this
formation process evolves, it must be completed
before the onset of accelerated expansion some 4 billlion
years ago at cosmic time $t \sim 9.8$ Gy.

\subsection{Like-Sign-Charged Primordial Extremely \\
Massive Naked Singularities (PEMNSs) and\\
Accelerated Expansion: the EAU Model}

\noindent
A novel EAU model was suggested in \cite{FramptonPLB,FramptonMPLA} where
 dark energy is replaced by charged dark matter in the form of PEMNSs or charged
Primordial Extremely Massive Naked Singularities\footnote{In 
\cite{FramptonPLB,FramptonMPLA} the PEMNSs were called PEMBHs}. That discussion 
involved the new idea that at the largest cosmological distances, {\it e.g.} greater than 1 Gpc, the dominant force
is electromagnetism rather than gravitation. 
\bigskip

\noindent
The production mechanism for PBHs in general is not well understood, and for the
PEMNSs we shall make the assumption that they are formed
before the accelerated expansion begins at $t=t_{DE}\sim 9.8$ Gy, For the expansion before $t_{DE}$ we shall assume
that the $\Lambda CDM$ model is approximately accurate.

\bigskip

\noindent
The subsequent expansion in the charged dark matter cPEMBH model will in the future
depart markedly from the $\Lambda CDM$ case. We can regard this as advantageous
because the future fate of the universe in the conventional picture does have certain
distasteful features in terms of the extroverse, as we briefly
review.

\bigskip

\noindent
In the $\Lambda CDM$  model the introverse, or what is also called the visible 
universe, coincides with the extroverse at $t=t_{DE} \sim 9.8$ Gy with the common
radius
\begin{equation}
R_{EV}(t_{DE}) = R_{IV}(t_{DE}) =  39 Gly.
\label{tDE}
\end{equation}

\bigskip

\noindent
The introverse expansion is limited by the speed of light and its radius increases
from Eq. (\ref{tDE}) to 44 Gly at the present time $t=t_0$ and asymptotes to
\begin{equation}
R_{IV} (t \rightarrow \infty) \rightarrow 58 Gly
\label{RIVasymp}
\end{equation}

\bigskip

\noindent
The extroverse expansion is exponential and superluminal. Its radius increases
from its value 39 Gly in Eq. (\ref{tDE}) to 52 Gly at the present time $t=t_0$ and grows without limit
so that after a trillion years it attains the extremely large value
\begin{equation}
R_{EV} (t  = 1 Ty) = 9.7 \times 10^{32} Gly.
\label{REVtrillion}
\end{equation}

\bigskip

\noindent
This future for the $\Lambda CDM$ scenario seems distasteful because the
introverse becomes of ever decreasing, and eventually vanishing, significance,
relative to the extroverse.

\bigskip

\noindent
A possible formation mechanism of PEMNSs was provided
in \cite{Araya2022} where their common sign of electric charge, negative,
arises from preferential accretion of electrons relative to protons. This
formation mechanism is not well understood
\footnote{Electrically neutral PEMBHS were first considered, with a different acronym SLABs,  in \cite{Carr2021}.}
so to create a cosmological model we shall for simplicity assume that
the PEMNSs are all formed before $t=t_{DE} \sim 9.8$ Gy and
thereafter the
Friedmann equation ignoring radiation, is
\begin{equation}
\left( \frac{\dot{a}}{a} \right)^2= \frac{\Lambda(t)}{3} + \frac {8\pi G}{3} \rho_{matter}
\label{Friedmann}
\end{equation}
where $\Lambda(t)$ is the cosmological constant generated by the Coulomb
repulsion between the PEMNSs.
From Eq.(\ref{Friedmann}), with $a(t_0) = 1$ and constant $\Lambda(t)\equiv \Lambda_0$, we
would predict that asymptotically in the future
\begin{equation}
a(t \rightarrow \infty) \sim exp \left( \sqrt{ \frac{\Lambda_0}{3}} (t-t_0) \right)
\label{exponential}
\end{equation}

\bigskip

\noindent
However, in the case of charged dark matter, with no dark energy, we must
re-write Eq.({\ref{Friedmann}) as
\begin{equation}
\left( \frac{\dot{a}}{a} \right)^2= \frac {8\pi G}{3} \rho_{cPEMNSs}+ \frac {8\pi G}{3} \rho_{matter}
\label{FriedmannPrime}
\end{equation}
in which
\begin{equation}
\rho_{matter} (t)  = \frac{\rho_{matter} (t_0)}{a(t)^3}
\label{rhomatter}
\end{equation}
where matter includes both normal matter and the uncharged dark matter.

\noindent
Of special interest in the present discussion is the expected future behaviour
of the charged dark matter
\begin{equation}
\rho_{PEMNSs} (t) = \frac{\rho_{PEMNSs} (t_0)}{a(t)^3}
\label{rhocPEMBHs}
\end{equation}
so that comparison of Eq.(\ref{Friedmann}) and Eq.(\ref{FriedmannPrime}) suggests
that the cosmological constant is predicted to decrease from its present value.
More specifically, we find
that asymptotically  the scale factor will behave as if matter-dominated
and the cosmological constant will decrease at large future times as a power
\begin{equation}
a(t\rightarrow \infty) \sim t^{\frac{2}{3}} ~~~~~~ \Lambda(t \rightarrow \infty) \sim t^{-2}.
\label{scale}
\end{equation} 

\bigskip

\noindent
so that a trillion years in the future $\Lambda(t)$ will have decreased
by some four orders of magnitude relative to $\Lambda(t_0)$. See
Table 1.

\begin{table}[t]
\caption{COSMOLOGICAL ``CONSTANT".}
\begin{center}
\begin{tabular}{||c|c||}
\hline
\hline
time & $\Lambda(t)$ \\
\hline
\hline
$t_0$ & $(2.0meV)^4$ \\
\hline
\hline
$t_0+10Gy$ & $(1.0 meV)^4$ \\
\hline
\hline
$t_0+100Gy$ & $(700 \mu eV)^4$ \\
\hline
\hline
$t_0+1Ty$ & $(230 \mu eV)^4$ \\
\hline
\hline
$t_0+1Py$ & $(7.4\mu eV)^4$ \\
\hline
\hline
\end{tabular}
\end{center}
\end{table}

\bigskip

\noindent
According to the $\Lambda CDM$ model, we live at a special time in
cosmic history because of the density coincidence between dark matter
and dark energy.
In the present case where charged dark matter replaces dark energy, the present
era is also special because the accelerated expansion,
discovered in 1998, is a temporary phenomenon centred around the present time.
Acceleration 
began about 4 Gy ago at $t_{DE}= 9.8Gy = t_0-4 Gy$.
This behaviour will disappear in a few more billion
years. The value of the cosmological constant is predicted to fall
like $a(t)^{-2}$ so that, when $t \sim \sqrt{2} t_0 \sim 19.5 Gy
\sim t_0 +4.7 Gy$, the value of
$\Lambda(t)$ will be one half of its present value, $\Lambda(t_0)$. 
As discussed in \cite{FramptonMPLA}, the equation of state associated 
with $\Lambda$ is predicted to be
extremely close to $\omega = -1$, so close that measuring the
difference seems impracticable.

\bigskip

\noindent
Let us  discuss the future time evolution of the introverse and
extroverse in the case of charged dark matter.  For the introverse,
nothing changes from the $\Lambda CDM$, and after a trillion years, the introverse radius
will be at its asymptotic value $R_{IV}=58 Gly$, as stated in Eq.(\ref{RIVasymp}).
By contrast, the future for the extroverse
is very different for charged dark matter. 
WIth the growth $a(t) \propto t^{\frac{2}{3}}$ we find
that the radius of the extroverse at $t=1$ Ty is 
\begin{equation}
R_{EV}(t=1Ty) \sim 900 Gly
\label{REVnew}
\end{equation}

\bigskip

\noindent
to be compared with the corresponding huge value
$9.7 \times 10^{32}$ Gly predicted by the $\Lambda CDM$ model, quoted in
Eq.(\ref{REVtrillion}) above. Eq.(\ref{REVnew}) means that if there still exist humans
in the Solar System, or at least in the Milky Way, their view
of the distant universe will include many billions of galaxies.

\bigskip

\noindent
In the $\Lambda CDM$ case, a hypothetical observational cosmologist, one trillion years
in the future,
could observe only the Milky Way and objects which are gravitationally bound to it, so that cosmology could become
an extinct science.
In the case of charged dark matter, for comparison, the time dependence
will allow about 180 billion out of a present trillion galaxies to
remain observable at $t=1Ty$ so that the view of the 
universe at that distant future time will look quite similar to 
the view at the present.

\bigskip

\noindent
The distinct physics advantage of charged dark matter is that it avoids the 
idea of an unknown repulsive gravity inherent in ''dark energy". Electromagnetism
provides the only known long-range repulsion so it is more attractive
to adopt it as the explanation for the accelerating universe.
A second advantage of charged dark matter 
is that it provides a conducive environment for cosmology, a trillion years in the
future.

\newpage

\section{Possible Support for the EAU Model from\\
the Amaterasu Cosmic Ray}

\noindent
Particle theory deals with very tiny particles which are typically smaller
than an atomic nucleus of size $10^{-15}$ m and therefore at least fifteen orders of magnitude
below the scales familiar to us. It treats objects far smaller than anything we can see with the
naked eye.
Theoretical cosmology, by contrast, deals with very large objects which are typically larger than the Milky Way galaxy of size $10^{23}$ m and hence in excess of twenty-three orders of magnitude larger than familiar scales. It considers objects so huge that they stretch the powers of our human imagination.

\bigskip

\noindent
An outsider could reasonable surmise that physicists who research particle theory
form an entirely separate group from the physicists who research theoretical cosmology because the two groups study scales which differ by over thirty-eight orders of magnitude.
However, it has been known for many decades that this surmise is mistaken because when
we consider the early universe the temperature can be so high that subnuclear particles are inevitably produced.
This fusion of the two research fields is sometimes displayed on an Ouroboros diagram, and the small-large connection has been very successfully exploited for over half a century.

\bigskip

\noindent
In the present section, we hope to convince the reader of the claim that a small (proton)-large (Local Void) fusion can exist even at the present time. The claim is based on the recent observation of a super-GKZ cosmic ray, called Amaterasu, which provides us
with a type of paradox whose resolution frequently results in a significant increase in human knowledge.

\bigskip

\noindent
Historically, the most important theoretical result for ultra high energy cosmic rays is the GKZ bound
\cite{Greisen:1966jv,Zatsepin:1966jv} that, to traverse the CMB, the energy is
bounded by
\begin{equation} 
E < 50 EeV
\label{GKZ}
\end{equation}
Observationally, over the years since \cite{Greisen:1966jv,Zatsepin:1966jv} the fortunes of the bound have ebbed and flowed but, at the present time, the cut off in Eq.(\ref{GKZ}) is very well established with only a few rare outliers exhibiting super-GKZ behaviour.
The Amaterasu's energy is $E=240$ EeV, the third largest ever recorded after
previous super-GKZ cosmic rays with $320$ EeV (1991) and $E=280$ EeV (2001).

\bigskip

\noindent
What makes the Amaterasu particle\cite{abbasi} doubly
interesting is that not only is it super-GKZ, but the direction tracks back to
the Local Void which contains no galaxies and therefore,  it was thought, 
no source\footnote{To allay all possible concerns that the primary direction used in \cite{abbasi} might be distorted by foreground
effects, we found the excellent review by Anchoroqui\cite{Anchordoqui:2018qom} to be convincing.}.

\bigskip

\noindent
The authors of \cite{abbasi}, however, restricted their attention to the
$\Lambda CDM$ model, without
considering the recently proposed EAU model
\cite{FramptonPLB,FramptonMPLA}. The latter forgoes the century old assumption that gravitation dominates electromagnetism at all length scales greater
than that characterising molecules, as tacitly assumed in a paper \cite{Einstein}
by the discoverer of relativity.  We shall argue in the present section that
the EAU model provides a natural resolution of the Amaterasu paradox.

\bigskip

\noindent
In the EAU model, all the dark matter is composed of Primordial Black Holes (PBHs)
with that in galaxies and clusters being Primordial Intermediate Mass Black Holes (PIMBHs), while at galactic centres there are Primordial Supermassive Black Holes (PSMBHs).
All of these PBHs are electrically neutral like the stars and planets. Only
Primordial Extremely Massive Naked Singularities (PEMNSs), with masses in excess of a trillion solar masses have negative\footnote{Note that if all the PEMNSs had, instead, a
positive charge our discussion of accelerated expansion would go through.}   electric charge with an overall charge asymmetry, relative to the
totatlty of the proton or electron charges, of
about one in a billion billion.

\bigskip

\noindent
Structure formation in galaxies and clusters, including
the Local Void, is due only to gravitational forces. On the other hand, the structure formation regarding PEMNSs is due to electromagnetic forces, and the two results regarding voids are expected to be quite different. In particular, what is the Local Void in terms of galaxies expected to contain PEMNSs and their electric charge can underly the origin of the Amaterasu cosmic ray.

\bigskip

\noindent
Consider a Primordial Extremely Massive Naked Singularity (PEMNS)
with mass $M_{PEMNS}=10^{12}M_{\odot}$ and negative electric
charge $q_{PEMNS}=-10^{32}$ Coulombs at a distance 1Mpc from the
Earth. Consider also a proton $p$ approximately at  rest, a candidate for
the Amaterasu primary, at a distance x metres behind the Earth and precisely 
aligned with the PEMNS and the Earth. To be justified {\it a posteriori} we
assume that x metres $<<$ 1 Mpc.

\bigskip

\noindent
The Coulomb attraction between PEMNS and $p$ is given by
\begin{equation}
F = \frac{ k_e q_{PEMNS} q_{\bar{p}}}{r^2}
\end{equation}
where the electric force constant is $k_e = 9 \times 10^9 N.m^2/C^2$.
Using $1Mpc = 3 \times 10^{22}m$ and proton charge
$+1.6\times 10^{-19}$ Coulombs gives an attractive electric force
which is approximately constant if x is sufficiently small
\begin{equation}
F= 1.6 \times 10^{-22} N
\end{equation}
in Newtons $N \equiv kg.m/s^2$. Inserting the proton mass
$m(p) = m_0 = 1.6 \times 10^{-19}$ kg the initial acceleration is
\begin{equation}
a_i = a(\beta_i=0) = \frac{F}{m_0} = 1.0 \times 10^5 m/s^2.
\label{initial}
\end{equation} 

\bigskip

\noindent
The required BKZ final relativistic velocity $\beta_f = v_f/c$ is given by
\begin{equation}
\frac{E_f}{m_0} = \frac{1}{\sqrt{1-\beta_f^2}} = \frac{2.4\times10^{20} eV}{938\times10^6 eV} = 2.56 \times 10^{11},
\end{equation}
so that
\begin{equation}
\beta_f^2 = 1 - 1.52 \times 10^{-23}.
\end{equation}

\noindent
For the relativistic acceleration of $\beta = v/c$ from
$\beta_i =0$ to $\beta_f = \sqrt{1-1.52 \times 10^{-23}}$
we may use the integral
\begin{equation}
\int \frac{dx}{\sqrt{1-x^2}} = \sin^{-1} x.
\label{integral}
\end{equation}

\bigskip

\noindent
We now integrate the motion from rest at time $t=t_i$ to reaching
energy $2.4 \times 10^{20}$ eV at time $t=t_f$ using the acceleration
\begin{equation}
\frac{d^2 s}{d t^2} =  c \frac{d \beta}{dt} = \frac{F}{m(\beta)} = \frac{F}{m_0} \sqrt{1-\beta^2} =
a_i \sqrt{1-\beta^2}
\end{equation}
with the initial acceleration $a_i = 10^5 m/s^2$ (see Eq.(\ref{initial})) and $c=3\times 10^8m/s$.

\bigskip

\noindent
Using the integral in Eq.(\ref{integral}) now gives the result
\begin{equation}
\beta_f = \sin \left[ \frac{t_f - t_i}{3000s} \right].
\label{time}
\end{equation}

\bigskip

\noindent
Since $\beta_f < 1$, we deduce from Eq.(\ref{time}) 
that $(t_f-t_i) < 3000s$ which implies that the initial 
at-rest proton must be less than $10^9$ km
from the Earth which is well within the Solar System,
actually within the orbit of Saturn.

\bigskip

\noindent
We emphasise that this requires precise alignment of the Amaterasu primary
with the PEMNS-to-Earth direction and this is expected only extremely
rarely. Nevertheless, our work does suggest that the Amaterasu cosmic ray
which hit the Earth's atmosphere on in May 2021 may remarkably shed light 
on the theory of the visible universe at the highest length scales.

\bigskip

\noindent
Cosmic rays have historically had a major r\^{o}le in particle
physics, such as the original discoveries of the positron,
the pion and many other hadrons. The Amaterasu cosmic ray is only one event but
it is an extraordinary one, as one of the three most energetic
cosmic rays ever recorded and the only one of those three pointing
back to the Local Void where, according to the
$\Lambda CDM$ model, there is no obvious source.

\bigskip

\noindent
We have discussed a possible
explanation for the Amaterasu particle where the source is a PEMNS residing
in the Local Void which locally accelerates a proton primary.
If our discussion is correct, this single cosmic ray has helped determine the correct choice
of theoretical cosmological model.

\newpage

\section{Discussion}

\noindent
We may have engaged in idle speculation in this paper but we are unaware
of any fatal flaw. We have replaced the conventional make up
for the slices of the universe's energy pie (5\% normal matter; 25\% dark matter;  70\% dark energy) with a similar but crucially changed version(5\% normal matter; 25\% dark matter;  70\% charged dark matter).

\bigskip

\noindent
The name dark energy was coined by Turner\cite{Turner} in 1998 shortly
after the announcement of accelerated expansion\cite{Perlmutter, Riess}.
An outsider familiar with $E=Mc^2$ might guess that dark energy
and matter are equivalent. If our model is correct, she would
be correct although it has nothing to do with $E=mc^2$. Charged 
dark matter replaces dark energy, an ill-chosen name because it 
suggested that there exists an additional component in the Universe.

\bigskip

\noindent
In the previous section, we argued that the unusual properties of the Amaterasu
cosmic ray reported in November 2023 could provide support for the
EAU model. More recently in April 2024, news \cite{DESI} from the Dark Energy 
Spectroscopic Instrument (DESI) at Kitt Peak in Arizona, USA, gave a
preliminary indication that the cosmological constant $\Lambda(t)$ 
is not constant but diminishing with time, as suggested by our Eq.(\ref{scale}),
and by our Table 1, thus providing a second possible support for the EAU model.

\bigskip

\noindent
Other supporting evidence could appear in the foreseeable future
from the James Webb Space Telescope (JWST) which might shed
light on the formation of PBHs in the early universe, also from the
Vera C. Rubin Observatory in Chile which will study long duration microlensing light 
curves  which could provide evidence for the existence of PIMBHs
inside the Milky Way.

\bigskip

\noindent
It will be interesting to learn how these and other observations may
support the idea that the observed cosmic acceleration is caused
by charged dark matter.

\newpage

\section*{Acknowledgement}

\noindent
We wish to thank
the editors of {\it Entropy} magazine for coming up with the idea of this
eightieth anniversary Festschrift, and to thank all of the other contributors for writing and
submitting their stimulating papers.

\end{document}